\documentclass[%
preprint,
 amsmath,amssymb,
prb,
]{revtex4-1}

\usepackage{graphicx}
\usepackage{dcolumn}
\usepackage{bm}

\usepackage{color}

\begin{document}

\title{Nonlinear Coupling of Linearly Uncoupled Resonators}

\author{M. Menotti}%
 \affiliation{Xanadu, 372 Richmond St. W, Toronto, ON, M5V 1X6, Canada}
 \author{B. Morrison}
   \affiliation{Xanadu, 372 Richmond St. W, Toronto, ON, M5V 1X6, Canada}
  \author{K. Tan}
   \affiliation{Xanadu, 372 Richmond St. W, Toronto, ON, M5V 1X6, Canada}
  \author{J.E. Sipe}
  \affiliation{Department of Physics, University of Toronto, 60 St. George St., Toronto, ON, M5S 1A7, Canada}
\author{Z. Vernon}
  \affiliation{Xanadu, 372 Richmond St. W, Toronto, ON, M5V 1X6, Canada}
\author{M. Liscidini}
 \email{marco.liscidini@unipv.it}
\affiliation{Impact Centre, University of Toronto, 411-112 College St., Toronto, ON, M5G 1L6, Canada}%
\affiliation{Dipartimento di Fisica, Universit\`a degli studi di Pavia, Via Bassi 6, 27100 Pavia, Italy}

\begin{abstract}
We demonstrate a system composed of two resonators that are coupled solely through a nonlinear interaction, and where the  linear properties of each resonator can be controlled locally. We show that this class of dynamical systems has peculiar properties with important consequences for the study of classical and quantum  nonlinear optical phenomena. As an example we discuss the case of dual-pump spontaneous four-wave mixing.
\end{abstract}

\maketitle

Many nonlinear phenomena involve a nonlinear coupling of the normal modes that characterize the system dynamics in the linear limit. For example, \textcolor{black}{counterpropagating modes can be nonlinearly coupled by the Kerr effect in Sagnac resonators \cite{kaplan81},} and thermal expansion in a solid is associated with a nonlinear coupling of phonons. In these examples the normal modes involve energy distributed throughout space. This feature of normal modes holds even in artificially structured materials. In optical physics, Gentry et al. \cite{gentry2014,popovic2015} have investigated four-wave-mixing (FWM) in a system composed of two or more coupled ring resonators, where nonlinear optical interactions couple the super-modes of the ``photonic molecule'' built from the resonators. A similar description holds for the nonlinear optics of a system of photonic crystal cavities.  \cite{azzini2013}And while nonlinear optical interactions can occur in more complex systems, including FWM in photonic crystal structures (PhC) \cite{Xiong:11} or coupled-resonator-optical-waveguides (CROWs) \cite{davanco2012}, in all these examples the normal modes of the system are extended over the full size of the structure.

An alternate scenario would involve nonlinear coupling between normal modes with energy localized in space. This is unusual. In the phonon example, it would arise if, for some reason, the Einstein model gave the correct description of lattice vibrations at the linear level rather than the Debye model; phonon-phonon interactions would then involve normal modes localized at lattice sites. Here linear group velocities would vanish, and any motion of excitations through the system would rely on the nonlinear interaction; the way in which the density of states could be modified by locally changing the properties of the system would be easily predicted and understood, and a simple ``local'' picture, rather than the more complicated ``wavevector'' or ``super-mode'' perspective, would be the natural one for system design. 

Since the coupling of energy in neighboring regions of space is usually more effectively done by linear interactions than by nonlinear interactions, such a situation does not typically occur. This is particularly true in traditional nonlinear optics, where nonlinear effects are usually so weak that their very observation requires intense powers  \cite{boyd,squeezing2016}. Even in integrated optical structures, where classical and quantum nonlinear optics have been demonstrated recently  at low input powers, \cite{Absil:00,Turner:08,ferrera2008,Clemmen:09,Li:2016aa,Razzari:2009aa,Levy:2009aa} in the systems investigated to date the linear coupling between excitations in different regions of space still dominates the nonlinear coupling.

\begin{figure}[b]
\centering
\includegraphics[width=0.45\textwidth]{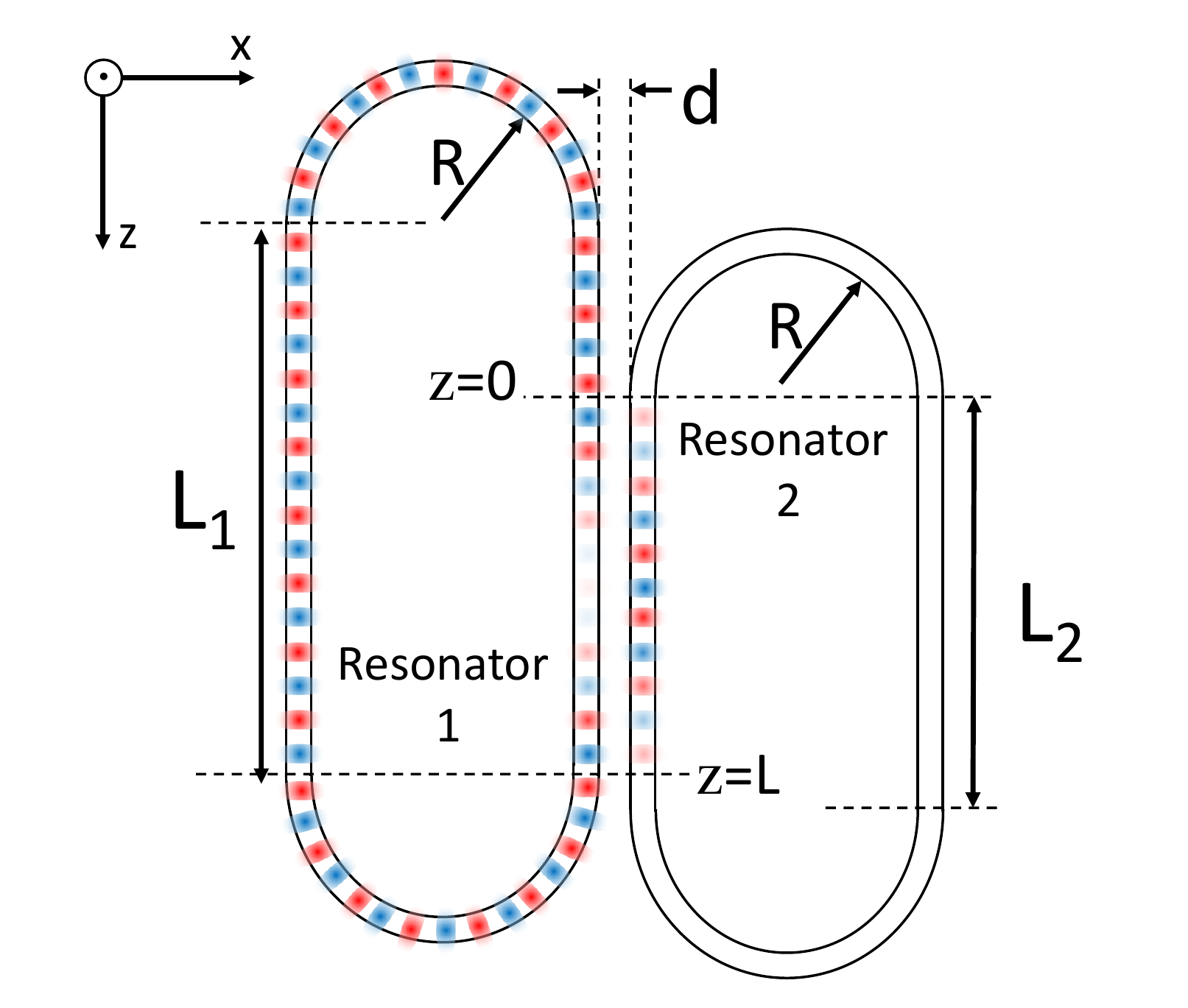}
\caption{Structure composed of two linearly uncoupled resonators.}
\label{fig:zing}
\end{figure}

Yet we show in this letter that by proper design one can arrange resonators that are \emph{linearly uncoupled}, in that normal modes are associated with one resonator or the other, but \emph{nonlinearly coupled}. We stress that this is not just because any effective linear coupling vanishes due to the lack of shared resonant frequencies \cite{Ou93}. Indeed, it survives even when the resonators have common resonances, and its most interesting consequences will be exhibited in that limit.

\textcolor{black}{As an example that highlights the significance of this kind of structure in quantum nonlinear optics}, we consider two racetrack resonators, as shown in Fig.\ref{fig:zing}. For simplicity we assume that the resonators have the same waveguide cross section and bending radius $R$ , but we allow the lengths $L_1$ and $L_2$ of the indicated straight sections to be different. The two racetracks are side-by-side, separated by a distance $d$, forming a directional coupler (DC) with a length $L$ that depends on the displacement of one resonator from the other along $z$. In general the DC will allow light to pass from one resonator to the other, and two super-modes will result as the normal modes of the system, each with significant energy in both resonators. However, a suitable choice of the length $L$ and the distance $d$ can guarantee vanishing transmission through the coupler, yielding a high isolation across a relatively wide bandwidth \cite{yariv}. If this is done, then the energy of the normal modes will be essentially confined to one or the other  resonator. In Fig. \ref{fig:zing} we indicate the field amplitude in one of the modes, which is restricted to resonator 1 except in the region of the DC; we say this mode is ``associated'' with resonator 1. Within the bandwidth of the DC there will be a set of such modes associated with resonator 1, and a corresponding set of modes associated with resonator 2. This separation of the distribution of energy in the two sets of modes will arise even if the resonators are of identical size and share the same resonance frequencies. Importantly, (i) \emph{the two sets of resonant frequencies can be controlled independently}, either through a proper choice of $L_1$ and $L_2$ or by means of local tuning elements, such as electric heaters, and (ii) since some of the energy of every mode is in the DC, \emph{in general there will be a nonlinear coupling even between modes associated with different resonators}. 
\begin{figure}[b]
\centering
\includegraphics[width=0.45\textwidth]{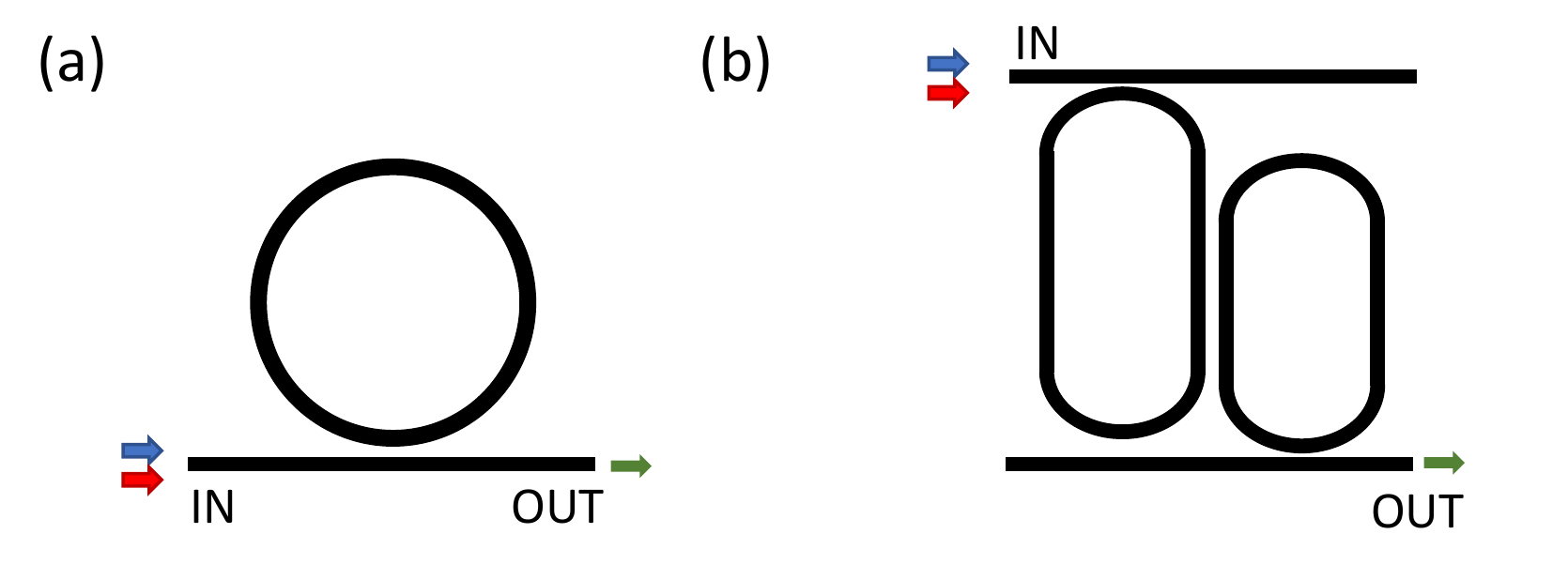}
\caption{Configuration of dual-pump SFWM  in  a single ring (a) and in two linearly uncoupled resonators (b).}
\label{fig:structure}
\end{figure}

\begin{figure}[t]
\centering
\includegraphics[width=0.48\textwidth]{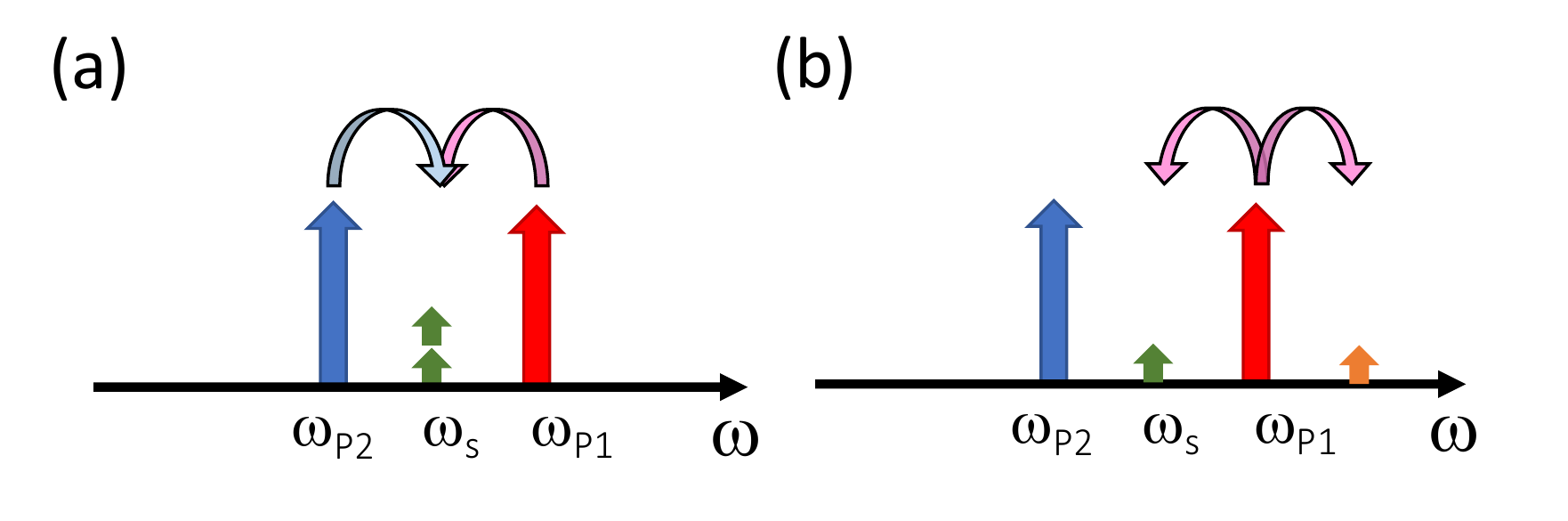}
\caption{(a) Dual-pump SFWM energy configuration; (b) Single-pump SFWM in the case of Pump 1.}
\label{fig:dual}
\end{figure}

To illustrate how these features can be put to use, consider the example of dual-pump spontaneous four-wave mixing (SFWM)\cite{kumar1984,silverstone2013,preble2015,vernon_arXiv}. In a usual scenario involving a single racetrack resonator (see Fig. \ref{fig:structure}(a)), pairs of photons can be generated at a frequency $\omega_S$ by photons from pump 1 at $\omega_{P1}$ and pump 2 at $\omega_{P2}$; the frequencies satisfy $\omega_{P1}$-$\omega_{S}$=$\omega_{S}$-$\omega_{P2}$, (see Fig. \ref{fig:dual} (a)). One of the main difficulties is the noise associated with unwanted photon pairs generated by pump 1 and pump 2 separately (see Fig. \ref{fig:dual} (b)). In these two processes one - and only one - of the photons is created with the same frequency of those generated via dual-pump SFWM, leading to a degradation of the quantum correlations of the generated light. \textcolor{black}{The noise given by these spurious photons is comparable to the signal associated with the generation of photon pairs at $\omega_S$. Indeed, assuming that all the resonances are equally spaced, thanks to proper dispersion engineering, and that the two pumps have the same intensity, the probability of generating a pair of photons at $\omega_S$ is the same as that of generating a non-degenerate photon pair with only one photon at $\omega_S$.} For this reason, these parasitic processes must be suppressed. Solutions based on dispersion engineering \cite{kim2017}, Bragg gratings \cite{agata2017}, and linearly coupled resonators have been proposed \cite{gentry2014,popovic2015}. 

Here we propose a strategy that  takes advantage of the local tunability of the two set of resonances of the structure shown in Fig.\ref{fig:zing}. We envision both pump fields in modes associated with resonator 1, and the signal generated in a mode associated with resonator 2, in the configuration depicted in Fig. \ref{fig:structure} (b). The two pumps are injected through the IN port, while the generated light exits from the OUT port. By choosing $L_1$ and $L_2$ - or by adjusting the effective length of the resonators by electric heaters after fabrication - we can ensure that energy conservation is satisfied \emph{only} for dual-pump SFWM, as we will see below. 

\begin{figure}[t]
\centering
\includegraphics[width=0.48\textwidth]{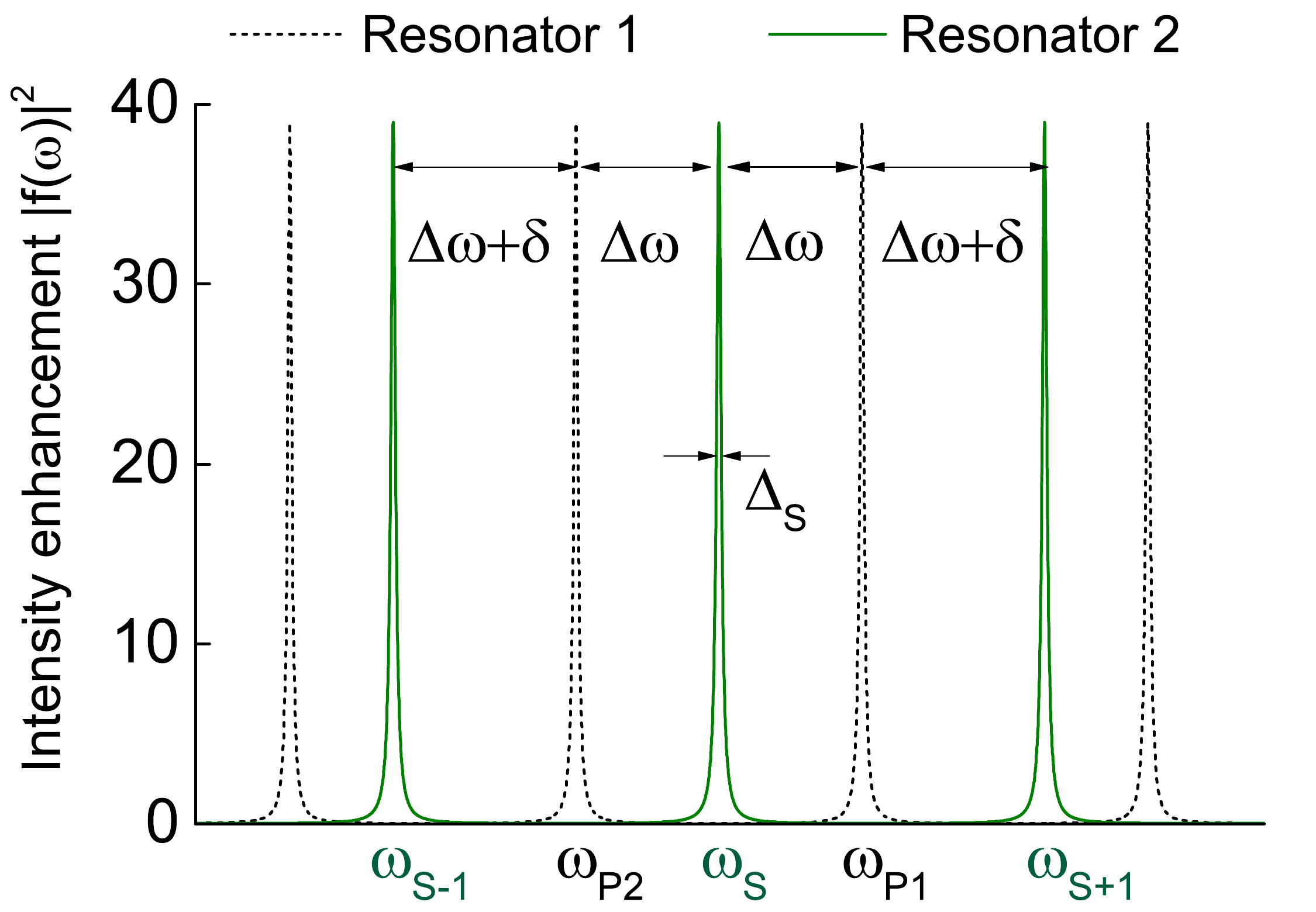}
\caption{(Color online) Intensity enhancement \textcolor{black}{$|f(\omega)|^2$} in the first resonator (dashed line) and in the second resonator (solid line) as a function of frequency.}
\label{fig:field_en}
\end{figure}

Note that just as the sets of resonances can be independently controlled, the coupling conditions and quality factors of each resonator can also be adjusted separately, either through the choice of the distance between the bus waveguide and the corresponding resonator, or by means of electrically tunable interferometric couplers\cite{chen07,preble2017}. Besides being present in the input channel, the intensity of both pumps will be mainly in resonator 1, and as well in the region of the DC. There the nonlinearity can act and generate light at $\omega_{S}$ , with intensity mainly in resonator 2 as well as being present in the output channel. In Fig. \ref{fig:field_en} we plot the \textcolor{black}{intensity} enhancement of the field associated with the resonances of the two linearly uncoupled resonators. The lengths $L_1$ and $L_2$ are chosen such that $\omega_{P1}$ and $\omega_{P2}$ are located symmetrically about $\omega_S$; yet it is clear that pairs of photons generated by single-pump SFWM from a pump at $\omega_{P1}$, say at $\omega_{S+1}$ and $\omega_S$, will not satisfy the energy conservation condition $\omega_{S+1}$+$\omega_{S}$=2$\omega_{P1}$, and thus these processes will be suppressed \textcolor{black}{by a factor $\Delta_s^2/(\Delta_s^2+\delta^2)$, with $\delta=2\omega_{P1}-\omega_{S+1}-\omega_{S}$ . Since $\delta$ can be comparable with the resonator free spectral range, for high-finesse resonators one can expect an improvement of the signal-to-noise (limited to the noise associated with the generation of spurious photons) of more than four orders of magnitude (see Supplemental Materials).} 

We now turn to the full calculation of the nonlinear coupling, taking into account the presence of the input and output channels and the coupling between them and the resonators. The nonlinear coupling strength is determined by the overlap integral \cite{helt2012}
\begin{eqnarray}\nonumber
J(\omega_1,\omega_2,\omega_3,\omega_4)=\int\mathrm{d}\mathbf{r}\Gamma^{ijkl}_3(\mathbf{r})\left[
D^{i,IN}(\mathbf{r}, \omega_3)\right.\\
\left.\times D^{j,IN}(\mathbf{r}, \omega_4)
D^{k,OUT}(\mathbf{r}, \omega_1)
D^{l,OUT}(\mathbf{r}, \omega_2)
\right], \label{eq:integral_asy}
\end{eqnarray}
where $\mathbf{D}^{IN(OUT)}(\mathbf{r}, \omega)$ is the properly normalized asymptotic-in field entering from the IN (OUT) port \cite{liscidini2012}.

The overlap integral (\ref{eq:integral_asy}) acquires non-vanishing contributions solely from the integration over the DC, the only region common to all the resonant modes involved. Here one can express the total field as a linear combination of fields associated with the individual waveguides forming the DC. Using standard coupled mode theory \cite{yariv}, the asymptotic fields in each waveguide can be written as
\begin{equation}\label{fields}
\mathbf{D_N}^{IN(OUT)}(\mathbf{r}, \omega)=f(\omega)\mathbf{d}(x,y)A_N^{IN(OUT)}(z)e^{\imath(k(\omega)z-\omega t)},
\end{equation}
where $N=1,2$ indicates the DC waveguide, belonging to either the first or second resonator. Here $\mathbf{d}(x,y)$ is the properly normalized transverse displacement field profile (which, for simplicity, we assume independent of frequency), $k(\omega)$ is the mode wave vector, and $A^{IN(OUT}_N(z)$ is the slowly varying field amplitude. Finally, $f(\omega)$ takes into account the resonant field enhancement as shown in Fig. \ref{fig:field_en}. In general, $f(\omega)$ can be a quite complicated function, for in it is embedded the complex interplay between multiple interacting resonances. However, when the resonators are linearly uncoupled, $f(\omega)$ is determined only by the ring scattering losses and the coupling with the channel. Nearby any resonance $\omega_j$, it can be approximated with the Lorentzian function \cite{onodera2016}
\begin{equation}\label{field_enhancement}
f(\omega)\approx\sqrt{\frac{4Q_jv_{g,j}}{\mathcal{L}_j\omega_j}}\sqrt{\frac{Q_j}{Q_{c,j}}}\frac{\Delta_j/2}{(\omega-\omega_j)+i\Delta_j/2},
\end{equation}
where $\mathcal{L}_j$ is the total length of the corresponding resonator, $v_{g,j}$ is the group velocity in the waveguides, $Q_j$ is the loaded quality factor, $Q_{c,j}$ is the quality factor determined solely by the coupling with the channel, and $\Delta_j=\omega_j/Q_j$. 

The functions $A^{IN(OUT)}_N(z)$ are completely determined by the boundary condition and the DC length, which we take to be $L=m\pi/|\kappa|$ for $m\in\mathbb{N}$, with $\kappa$ the coupling constant between the two waveguides. \textcolor{black}{We assume that the Kerr effect does not affect the DC coupling efficiency, in line with experimental studies of photon pair generation in single-racetrack silicon resonators where no power dependence of the coupling efficiency between the waveguide and resonator was observed \cite{engin2013}  (See additional details and estimates in Supplemental Material.)}

Under these assumptions the two resonators are linearly uncoupled with
\begin{equation}\label{amplitudes}
\left\{ \begin{array}{ccc}
A_{1}^{OUT}(z) &=&\cos(|\kappa|z)\\
A_{2}^{OUT}(z) &=&i\frac{\kappa^{\ast}}{|\kappa|}\sin(|\kappa|z)\\
A_{1}^{IN}(z) &=&-i\frac{\kappa^{\ast}}{|\kappa|}\sin(|\kappa|z)\\
A_{2}^{IN}(z) &=&\cos(|\kappa|z)\\
\end{array}\right.
\end{equation}
Now using (\ref{fields}) and (\ref{amplitudes}) in (\ref{eq:integral_asy}), substituting $\Gamma^{ijkl}(\mathbf{r})= \Gamma^{ijkl}_3(x,y)=\chi^{ijkl}_3(x,y)/[\varepsilon_0^2n(x,y)^8]$, and assuming phase matching (or simply $L\ll\pi/|\Delta k|$, with $\Delta k=2k(\omega_S)-k(\omega_{P1})-k(\omega_{P2})$), we obtain
\begin{eqnarray}\label{joverlap_ZING}\nonumber
J(\omega_1,\omega_2,\omega_3,\omega_4)
&=&\frac{16v_g^2Q_{P}Q_{S}}{\mathcal{L}_1\mathcal{L}_2\omega_{S}\sqrt{\omega_{P1}\omega_{P2}}}\frac{Q_PQ_S}{Q_{c,P}Q_{c,S}}\frac{\bar{\chi_3}}{\bar{n}^4\mathcal{A}}\frac{L}{4}\\
&\times&\mathcal{F}(\omega_1,\omega_2,\omega_3,\omega_4),
\end{eqnarray}
where $\mathcal{A}$ is the effective area associated with the nonlinear process, $\bar{n}$ and $\bar{\chi}_3$ are respectively typical values for the refractive index and the nonlinear susceptibility, and for simplicity we take $Q_{P_1}\simeq Q_{P2}=Q_{P}$ and $v_{g,P1}\simeq v_{g,P2}\simeq v_{g,S}=v_g$.  Finally,
\begin{widetext}
\begin{eqnarray}
\mathcal{F}(\omega_1,\omega_2,\omega_3,\omega_4)=\frac{\Delta_{P1}/2}{(\omega_3-\omega_{P1})+i\Delta_{P1}/2}\frac{\Delta_{P2}/2}{(\omega_4-\omega_{P2})+i\Delta_{P2}/2}\frac{\Delta_S/2}{(\omega_1-\omega_S)+i\Delta_S/2}\frac{\Delta_S/2}{(\omega_2-\omega_S)+i\Delta_S/2}
\end{eqnarray}
\end{widetext}
describes the frequency dependence of the overall field enhancement.

It is interesting to compare \eqref{joverlap_ZING} to the strength $J_0(\omega_1,\omega_1,\omega_3,\omega_4)$ of the nonlinear interaction due to dual-pump SFWM in a single resonator (see Fig.\ref{fig:structure} (a)). For example, taking the resonator length $\mathcal{L}\simeq\mathcal{L}_1\simeq\mathcal{L}_2$, with equally spaced resonances having quality factor $Q\simeq Q_S\simeq Q_P$, one finds:
\begin{equation}
J(\omega_1,\omega_1,\omega_3,\omega_4)=J_0(\omega_1,\omega_1,\omega_3,\omega_4)\frac{L}{4\mathcal{L}},
\end{equation}
which shows that in our structure (Fig. \ref{fig:structure} (b)) the nonlinear interaction is reduced by a factor $L/(4\mathcal{L})$ from that of a single resonator. The attenuation comes from two independent effects: (i) for nonlinearly coupled resonators the interaction occurs only in the DC, while in a single ring it occurs over the entire resonator length; (ii) the slowly varying components of the pump and signal fields in the DC oscillate with opposite phases, leading to the factor $1/4$ in \eqref{joverlap_ZING}.
Yet this decrease in the interaction length comes with at least four major advantages: (i) \emph{Parasitic processes are suppressed}. This feature is inherent in our structure design, while additional strategies are required for a single resonator; most of the proposed solutions lead to a reduction of the field enhancement and thus of the nonlinear interaction strength;
(ii)  \emph{Waveguide dispersion engineering is not critical}. Unlike a single resonator, where   phase matching (i.e. $\Delta k=0$, including cross-phase modulation/self-phase modulation (XPM/SPM) effects) is required to ensure equally spaced resonances, this is not true here, for one only needs $|\Delta k|\ll\pi/L$ to operate at arbitrarily large quality factors. \cite{footnote} (iii) Assuming a tuning mechanism or a proper choice of the resonator lengths, the same device can virtually \emph{operate at any pump power}, for one can compensate any effect arising from XPM/SPM; (iv) Since the two resonators are ideally uncoupled, in a realistic device one can expect a \emph{pump rejection} of several tens of dB. \textcolor{black}{All these advantages are obtained \emph{simultaneously}, leading to a very flexible and powerful device for the control of nonlinear interactions.}


We now turn to the optimization of the nonlinear interaction strength, which is maximum when all the fields are on resonance. Assuming  quality factors limited by scattering losses, the value $J(\omega_S,\omega_S,\omega_{P_1},\omega_{P_2})$ is strongly determined by the mode volume and the coupling length $L$. On the one hand, high $Q$ and small $\mathcal{L}_{1,2}$ are desirable to maximize the field enhancement according to \eqref{field_enhancement}. On the other hand, the nonlinear interaction strength is directly proportional to the DC length $L$. These considerations lead to the natural choice of $\mathcal{L}_1\simeq\mathcal{L}_2=2(L+\pi R)$ with the optimal length $L_\mathrm{opt}=\pi R$.  \textcolor{black}{For a silicon-on-insulator device we can conservatively consider $R$ of 10-20 $\mu$ m and  Q of the order of $10^4-10^5$, which would guarantee a pair generation rate of about 1MHz with sub-mW pump powers \cite{engin2013,savanier16}.}

The example just discussed illustrates how the control and the enhancement of the nonlinear light-matter interaction requires the fulfillment of many different conditions simultaneously, e.g., field enhancement, energy conservation, phase matching, and the like.  Using linearly uncoupled resonators this can be simply done because of the local nature of the linear properties of the system. Indeed, sets of resonant frequencies can be identified with localized regions of space and therefore can be controlled independently.
It is clear that the same approach will lead to critical advantages in other classical and quantum nonlinear processes, either based on second- or third-order nonlinearities. For instance, in optical parametric oscillation (OPO), one will be able to work in either the normal or anomalous dispersion regime, at different pump power levels, and with the possibility of triggering OPO in different wavelength ranges. As another example, in single-photon frequency conversion via resonant four-wave-mixing Bragg scattering, one will be able to tune the resonance positions to suppress parasitic processes and selectively enhance only the desired wavelength conversion. \textcolor{black}{We also stress that the structure presented in Fig.1 is not the only way in which one can implement nonlinear coupling of linearly uncoupled resonators. For example, the DC in our structure could be replaced with an interferometric coupler\cite{chen07}.}

From a fundamental point of view, one can imagine more complex situations involving more than two nonlinearly coupled resonators.  Photonic molecules or crystals could be constructed, where the interaction between the constituent resonators would be \emph{exclusively} nonlinear. These systems would be characterized by complex nonlinear dynamics that are yet to be discovered and studied, and which could be controlled effortlessly by modifying the linear properties of each resonator independently.



%
\newpage

\section{Supplemental Material}
\subsection{Side-band process intensity in dual-pump SFWM}

In this section we want to estimate the strength of the side-band
processes in the dual-pump SFWM.

The number of generated photon pairs per pulse, is given by \cite{onodera2016}
\begin{equation}
|\beta|^{2}\approx\frac{|\alpha|^{4}(\hbar\omega_{s})^{2}}{\Delta T}\frac{9\pi^{3}}{2\varepsilon_{0}^{2}}\frac{1}{v_{g}^{4}}\int d\omega(2\omega_{s}-\omega)\omega|J(\omega,2\omega_{s}-\omega,\omega_{s},\omega_{s})|^{2},\label{eq:number_general}
\end{equation}
where the integral is performed on the second-signal resonance centred
at $\omega_{s2}$. Here $|\beta|^{2}$ is the average number of generated
pairs for a square pulse of duration $\Delta T$ and an average number
of photons $|\alpha|^{2}.$ 

The function $J(\omega,2\omega_{s}-\omega,\omega_{s},\omega_{s})$
contains the information related to the structure under consideration.
In the case of a ring resonator we have

\begin{equation}
|J(\omega,2\omega_{s}-\omega,\omega_{s},\omega_{s})|^{2}=\mathcal{K}\left(\frac{2}{1-\sigma}\right)^{4}\left(\frac{\left(\Delta/2\right)^{2}}{\left(\omega_{s2}-\omega\right)^{2}+\left(\Delta/2\right)^{2}}\right)\left(\frac{\left(\Delta/2\right)^{2}}{\left(\omega_{s2}+\delta-\omega\right)^{2}+\left(\Delta/2\right)^{2}}\right),\label{eq:overlap_detuning}
\end{equation}
where $\Delta$ is the full-width at half maximum of the resonance,
$\delta=2\omega_{s}-\omega_{p}-\omega_{s2}$, and $\mathcal{K}$ is
a constant that is related to the nonlinear parameter of the ridge
waveguide.

Inserting (\ref{eq:overlap_detuning}) in (\ref{eq:number_general})
and performing the change of variable $\omega_{s2}-\omega=u$, we
have 

\begin{eqnarray*}
|\beta|^{2}&\approx&\frac{|\alpha|^{4}(\hbar\omega_{s})^{2}}{\Delta T}\frac{9\pi^{3}}{2\varepsilon_{0}^{2}}\frac{\mathcal{K}}{v_{g}^{4}}\left(\frac{2}{1-\sigma}\right)^{4}\\
&\times&\int_{-\Omega}^{+\Omega}du(2\omega_{s}-\omega_{s2}+u)(\omega_{2s}-u)\left(\frac{\left(\Delta/2\right)^{2}}{u^{2}+\left(\Delta/2\right)^{2}}\right)\left(\frac{\left(\Delta/2\right)^{2}}{\left(u+\delta\right)^{2}+\left(\Delta/2\right)^{2}}\right),
\end{eqnarray*}
and since the product of the two Lorentzians is non-zero only if $u$
is close to zero and $\delta\leq\Delta/2$, using $\Delta\ll\omega_{s}$
we can also write 

\[
|\beta|^{2}\approx\frac{|\alpha|^{4}(\hbar\omega_{s})^{2}}{\Delta T}\frac{9\pi^{3}}{2\varepsilon_{0}^{2}}\frac{\mathcal{K}}{v_{g}^{4}}\left(\frac{2}{1-\sigma}\right)^{4}\omega_{2s}\omega_{p}\int_{-\infty}^{+\infty}du\left(\frac{\left(\Delta/2\right)^{2}}{u^{2}+\left(\Delta/2\right)^{2}}\right)\left(\frac{\left(\Delta/2\right)^{2}}{\left(u+\delta\right)^{2}+\left(\Delta/2\right)^{2}}\right).
\]
For $\delta=0$ (i.e., equally spaced resonances) we have 

\begin{equation}
|\beta_{0}|^{2}\approx\frac{|\alpha|^{4}(\hbar\omega_{s})^{2}}{\Delta T}\frac{9\pi^{3}}{2\varepsilon_{0}^{2}}\frac{\mathcal{K}}{v_{g}^{4}}\left(\frac{2}{1-\sigma}\right)^{4}\omega_{2s}\omega_{p}\frac{\pi}{4}\Delta,\label{eq:on_res}
\end{equation}
and otherwise

\begin{equation}
|\beta|^{2}\approx\frac{|\alpha|^{4}(\hbar\omega_{s})^{2}}{\Delta T}\frac{9\pi^{3}}{2\varepsilon_{0}^{2}}\frac{\mathcal{K}}{v_{g}^{4}}\left(\frac{2}{1-\sigma}\right)^{4}\omega_{2s}\omega_{p}\frac{\pi}{4}\frac{\Delta^{3}}{\delta^{2}+\Delta^{2}}.\label{eq:off_res}
\end{equation}
Then comparing (\ref{eq:on_res}) and (\ref{eq:off_res}) we identify
the attenuation due to the resonance detuning $\delta,$

\[
\frac{|\beta|^{2}}{|\beta_{0}|^{2}}=\frac{\Delta^{2}}{\delta^{2}+\Delta^{2}}.
\]

\subsection{Role of the Kerr effect in the coupling region}

Here we consider the role of the Kerr effect in the coupling region.
The importance of the Kerr effect on directional coupling has been
well investigated in the literature. There are two main effects related
to the change of the refractive index in the presence of an intense
optical field: (i) a change in the field distribution, which ultimately
controls the coupling coefficient $\kappa$ between the two waveguides,
and (ii) a change of the mode propagation constant. 

The first effect is related to self-focusing, and it has been studied
in fiber-based systems and in LiNbO$_{3}$ directional couplers, e.g. Ref. [\onlinecite{feit1988}]. It occurs at very intense optical fields, when the refractive
index variation induced by the Kerr effect is large enough to modify
the confinement of light in the structure; it has never been observed
on a silicon-on-insulator platform, where two-photon absorption limits
the available circulating power. 

The second effect, which has been investigated in detail by Stegeman
et al. in Ref. [\onlinecite{stegeman}], is definitely more important, as small variations
in the mode propagation constant can become relevant if the coupling
coefficient $\kappa$ is small. In this case, due to the difference
$\Delta\beta$ of the mode propagation constants in the two waveguides
forming the DC, the coupling efficiency is reduced, with $\kappa\rightarrow\eta$ (see Ref. [\onlinecite{yariv}]), where

\begin{equation}
\eta=\frac{|\kappa|^{2}}{|\kappa|^{2}+(\Delta\beta)^{2}}sin^{2}\left[|\kappa|L\sqrt{1+\left(\frac{\Delta\beta}{2|\kappa|}\right)^{2}}\right],
\end{equation}
and $L$ is the DC length. 

In our example we investigate the use of linearly uncoupled resonators
for the generation of degenerate photon pairs by spontaneous four-wave
mixing, and since we want to work in a regime in which $L\sim1/\kappa$,
the Kerr effect will be negligible as long as $\Delta\beta L\ll1$.
We can estimate $\Delta\beta$ from the input power $P_{in}$ in the
ring and the nonlinear parameter $\gamma$ of the waveguide, 

\begin{equation}
\Delta\beta=\frac{2\pi}{\lambda}\Delta n_{Kerr}\approx\gamma P_{in}\mathcal{F},
\end{equation}
where $\mathcal{F}$ is the resonator finesse, and in our structure
the optimal coupling length is $L_{opt}=\pi R$, 
with $R$ the bending radius of the race track. The condition $\Delta\beta L\ll1$
then gives 

\begin{equation}
\gamma P_{in}\mathcal{\mathcal{F}}\pi R\ll1,
\end{equation}
or alternatively 

\begin{equation}
\gamma P_{in}\frac{\lambda Q}{4n_{g}}\ll1\label{eq:onlyQ}
\end{equation}
since optimally $\mathcal{F=\lambda}Q/(4\pi n_{g}R)$, with $n_{g}$
the group index and a total resonator length of $4\pi R$. 

For a typical silicon-on-insulator device, we can take $\gamma\approx200\quad W^{-1}m^{-1}$,
pump powers of 5 mW, $\lambda=1500\quad nm$, $n_{g}\approx3$ and
$Q=50000$, which leads to 

\[
\left(2\times10^{2}\quad W^{-1}m^{-1}\right)\left(5\times10^{-3}W\right)\frac{\left(1.5\times10^{-6}m\right)\left(5\times10^{4}\right)}{12}\approx6.2\times10^{-3},
\]
which confirms our neglect of the Kerr effect on the DC operation,
and is in agreement with previous work on photon pair generation in
a single silicon racetrack resonator, where no power dependence of
the coupling coefficient between the waveguide and resonator was observed
\cite{engin2013}. 

Nonetheless, if the pump power were large enough to achieve optical
parametric oscillation in a SiN integrated device, the Kerr effect
could begin to modify the DC coupling efficiency. Taking $\gamma\approx1\quad W^{-1}m^{-1}$,
pump powers of 500 mW, $\lambda=1500\quad nm$, 2 and $Q=10^{6}$,
we find 

\[
\left(1W^{-1}m^{-1}\right)\left(5\times10^{-1}W\right)\frac{\left(1.5\times10^{-6}m\right)\left(10^{6}\right)}{8}\approx10^{-1}.
\]
But such pump powers are larger than we envision using in this work.

\end{document}